# Bethe-Salpeter Amplitudes and Density Correlations for Mesons with Wilson Fermions

Rajan Gupta, David Daniel, Jeffrey Grandy

*T-8, MS-B285, Los Alamos National Laboratory, Los Alamos, NM 87545*

We present an investigation of various gauge invariant definitions of the $q\overline{q}$ Bethe-Salpeter (BS) amplitude for mesons in lattice QCD, and compare them to the Coulomb and Landau gauge BS amplitudes. We show that the gauge invariant BS amplitude is considerably broadened by the use of "fat" gauge links (constructed by smearing the links of the original lattice). A qualitative demonstration of the Lorentz contraction of the pion wavefunction at non-zero momentum is given. We also calculate density-density correlations and discuss the limitations in extracting the charge radius of the pion from these observables. Lastly, the polarization dependence of the BS amplitude for the $\rho$ meson is exhibited, and we extract the relative admixture of $l=0$ and $l=2$ states showing that simple hydrogen-like wavefunctions provide a good fit.

## 1. Introduction – Bethe-Salpeter amplitudes

In the study of bound states in relativistic quantum field theory, commonly used tools are the Bethe-Salpeter (BS) amplitudes. These give the probability amplitude for finding some specified arrangement of fundamental components within the bound state. In this paper we study three different types of equal-time BS amplitudes for the pi and rho mesons and discuss whether one can extract an experimentally measured quantity such as the charge radius from such probes.

A gauge invariant definition of BSA for, say, a pion of momentum $\vec{p}$ is

$$\mathcal{A}_\pi(\vec{x}) = \langle 0|\overline{d}(\vec{0})\gamma_5 \mathcal{M}(\vec{0},\vec{x})u(\vec{x})|\pi(\vec{p})\rangle \qquad (1.1)$$

where $\mathcal{M}(\vec{0},\vec{x})$ is a path-ordered product of gauge links that joins points $\vec{x}$ and $\vec{0}$ and makes the amplitude gauge invariant. This amplitude is given by the following ratio of 2-point correlators:

$$\mathcal{A}_\pi(\vec{x},t) = \frac{\langle 0|\overline{d}(\vec{0};t)\gamma_5 \mathcal{M}(\vec{0},\vec{x};t)u(\vec{x};t)\,\overline{u}(\vec{y};0)\gamma_5 d(\vec{y};0)|0\rangle}{\langle 0|\overline{d}(\vec{0};t)\gamma_5 u(\vec{0};t)\,\overline{u}(\vec{y};0)\gamma_5 d(\vec{y};0)|0\rangle} \,. \qquad (1.2)$$

$\mathcal{M}$ may be defined in a number of ways. In this paper we use the APE smearing method [1] to construct "fat" gauge links and thus obtain a nonlocal average of paths for $\mathcal{M}$. We also consider the Coulomb gauge ($\mathcal{C}_\pi$) and Landau gauge ($\mathcal{L}_\pi$) BS amplitudes. These are obtained by transforming the quark propagators to Coulomb (Landau) gauge, and then calculating the ratio given in Eq. (1.2) with $\mathcal{M} = 1$, i.e. without including the links. The pion decay constant $f_\pi$ is related to $\mathcal{A}_\pi(0)$ and has already been calculated for this data set in Ref. [2]. Here we are only interested in the behavior of $\mathcal{A}_\pi(\vec{x})$ as a function of $\vec{x}$ so it is expedient to use the normalization $\mathcal{A}_\pi(0,t) = 1$.

Calculations of BS amplitudes in lattice QCD began with the 1985 study of Velikson and Weingarten [3]. Other calculations include the work of Chu et al. [4], Hecht and DeGrand [5], Duncan et al. [6], and Kilcup [7]. Most of this work has used the Coulomb or Landau gauge, though there is some work on simple gauge invariant prescriptions [4][5]. A preliminary version of the results presented in this paper have been reported earlier in Ref. [8]. Another similar probe of the hadronic structure are density-density correlations, as suggested in Refs. [9] [10] [4]. We discuss this method in section 5.

For gauge theories, the definition of BS amplitudes is complicated by the problem of gauge dependence. In QED the gauge dependence can be calculated and is typically $O(\alpha_{em})$ [11]. Therefore one is free to choose a gauge that makes calculations simple. This is why it is common to choose Coulomb gauge for QED calculations as it has the useful property of regulating infra-red divergences [12]. One does not expect such a simplification in a strongly coupled theory like QCD, instead to measure the distribution of constituents inside a hadron one needs a probe that takes into account the chromo-electric flux and $q\bar{q}$ pairs in addition to the valence quarks and antiquarks.

In the case of mesons we imagine that the valence $q$ and $\bar{q}$ are effectively confined by a flux tube. (In this picture the effect of additional $q\bar{q}$ pairs from the vacuum is to simply decrease the value of the string tension.) Since it is known from experiments that the glue carries a significant fraction of the hadron's momentum, different definitions of BSA will give very different answers depending on the overlap of the probe with the physical state. For this reason we characterize the "goodness" of a probe by the magnitude of the amplitude.

The virtue of Coulomb and Landau gauges is that they make color fields smooth by reducing the gauge fluctuations, so the fixed gauge BSA measures that part of the meson wavefunction corresponding to a smooth distribution of chromo-electric flux between the $q\bar{q}$. The principal disadvantage of calculations in these gauges is the poorly understood effect of Gribov copies. A review of the Gribov ambiguity in lattice calculations can be found in Ref. [13]. On the other hand it is obvious from Eq. (1.1) that in gauge invariant definitions each different choice of the connection $\mathcal{M}$ will give a different BSA. By using smeared links to construct the path joining $q\bar{q}$ we are using a certain linear combination of paths on the original lattice. There is, however, little control over the relative weights of the different paths, and in this sense the smearing process is not optimal. The usefulness of the smearing process lies in the fact that the overall thickness of the flux tube in the probe is controlled by the number of smearing steps and that the method is computationally simple.

It should be emphasized that while the BS amplitudes are simple to measure, there is, for light quarks, no demonstrated connection between a hadron's size extracted from them and the experimentally measured charge radius. What BS amplitudes do provide is a qualitative understanding of hadronic structure. This knowledge is useful for lattice studies in designing better probes, *i.e.* those that have a larger projection onto a given state. This technical improvement is crucial for improving the calculation of the spectrum and of matrix elements between hadronic states.

This paper is organized as follows. In Section 3 we present results for the gauge invariant BS amplitude as a function of the smearing size of the links and show that the amplitude becomes broader on using "fat" links to connect the $q\bar{q}$. We also give a qualitative demonstration of the Lorentz contraction of the wavefunction at non-zero momentum. A comparison between Coulomb gauge, Landau gauge and gauge-invariant amplitudes is made in Section 4. Calculation of density-density correlations is presented in Section 5 along with a discussion of why the charge radius measured in experiments is different from similar quantities that can be extracted from lattice data. In Section 6 we show that the BS amplitude for the $\rho$ meson depends on the polarization axis. We extract the relative admixture of $l = 0$ and $l = 2$ states from this data and find that simple hydrogen-like wavefunctions provide a good fit. Final conclusions are given in section 7.

## 2. Details of the lattices and propagators

The calculation is done using 35 background gauge configurations of size $16^3 \times 40$ at $\beta = 6.0$. The Wilson action quark propagators are calculated on doubled lattices ($16^3 \times 40 \to 16^3 \times 80$) using Wuppertal sources. The quark masses used are $\kappa = 0.154$ and $0.155$, corresponding to pions of mass 700 and 560 MeV respectively. Further details of lattice generation and propagator inversion are given in Ref. [2]. The main limitation of this calculation is that the spatial lattice size, $L = 16$, is not large enough to reliably extract the asymptotic behavior for some BS amplitudes and density-density correlations. We shall address this issue at appropriate places in the analysis.

We use a stochastic over-relaxed ($\omega = 2$) algorithm for gauge fixing to either Coulomb or Landau gauge. In the iterative process, the new gauge transformation at a given site is set to either $G_{new} = \Delta G * G_{old}$ or $G_{new} = \Delta G * \Delta G * G_{old}$ with equal probability. The change $\Delta G$ is the SU(3) matrix that maximizes Tr $(\Delta G \Sigma^{\dagger})$ where $\Sigma$ is the sum of links originating from that site. The convergence criteria used in both cases is $(3 - \text{Tr } \Delta G)/3 < 10^{-6}$. Operationally, we first fix the lattice to Landau gauge and then fix to Coulomb gauge as this saves computer time.

## 3. Gauge invariant BS amplitude with smearing

In an earlier calculation Chu *et al.* [4] investigated the simplest version of the gauge-invariant BS amplitude; for $\mathcal{M}(\vec{0}, \vec{x})$, they used the straight line path between points that lie along one of the lattice axes. They found that $\langle r^2 \rangle_\pi$ measured from the simplest gauge invariant BS amplitude is smaller than that obtained in either Coulomb or Landau gauge and that the latter estimates are 0.3–0.5 of the charge radius measured in experiments. Similar results have also been obtained by Hecht and DeGrand [5].

The poor quality of these earlier gauge invariant results is due to having used a straight line path to join the quark and anti-quark: such a probe has a poor overlap with the hadronic state because there is only a small probability amplitude for the gluon field to be so localized. The overlap can be improved significantly by using a smeared gluon field, and the resulting gauge-invariant BS amplitude is markedly enhanced.

To construct "fat" links we use the APE smearing method, which was first introduced to enhance the signal in glueball calculations [1]. In this method each link in the spatial direction $i$ is replaced by the sum

$$U_i^{(1)}(\vec{x}, \vec{x}+\hat{i}) = \mathcal{P}\left(U_i(\vec{x}, \vec{x}+\hat{i}) + \sum_{i=1}^{4} \mathcal{S}_i(\vec{x}, \vec{x}+\hat{i})\right) \tag{3.1}$$

where $\mathcal{S}_i$ are the four spatial staples shared by the link $U_i$, and the symbol $\mathcal{P}$ implies that the sum is projected back on to the group SU(3). As a result of this projection the normalization of the BS amplitude on different smeared levels is the same, so we can directly compare the results. The smeared link represents locally averaged gauge fields and has the same gauge freedom as the original link, *i.e.* under $U_i(x) \to g(x)U_i(x)g(x+i)$ the smeared link also transforms as $U_i^{(1)}(x) \to g(x)U_i^{(1)}(x)g(x+i)$. Note that the new lattice is of the same size as the original. One can iterate this smearing step as many times as necessary, using the effective fields at step $(n-1)$ to produce still "fatter" fields at step $n$; for example in the second smearing step the right hand side of Eq. (3.1) is constructed from smeared links produced in step one. A straight line path between the $q$ and the $\overline{q}$ that is made up of smeared links is in fact an average over a large number of paths on the original lattice. We specify the smearing level by a superscript on $\mathcal{A}$, which will be $0-6$ corresponding to the original links and six levels of smearing. We do not consider it appropriate to go beyond 6 levels of smearing on a lattice of size 16 with periodic boundary conditions.

It is worth mentioning that we had hoped to show that the chromoelectric flux is more spread out at small separations and gets confined to a flux tube at large separation by contrasting data at the different smearing levels. Unfortunately, the present data do not show any such effect.

We first study the behavior of the BS amplitude as a function of the time-slice $t$ from the source in order to determine how fast it converges. In Fig. 1a we show data for $\mathcal{A}_\pi^{(6)}$ for $t = 2, 5, 10$ and $15$, and similar data for the rho in Fig. 1b. The qualitative shape of the BS amplitude is similar for all $t$, though quantitatively it gets significantly broader, reaching an asymptotic value by about $t = 15$ for both the rho and the pion. (The difference between $t = 10$ and $t = 15$ is significant for the pion but not for the rho.) This separation is somewhat larger than $t = 10$ at which we find the onset of the plateau in the effective mass plots (which is taken as evidence that the correlator is dominated by the lightest state) as shown in Ref. [2] using the same set of lattices. The errors in the data are independent of $t$ for the pion and increase with $t$ for the rho. In Figs. 1a,b, we show the jackknife errors at all $x$ at $t = 2$ and for $x > 10$ at $t = 15$ as typical examples. We consider the $t = 15$ data best with respect to extracting the ground state and the statistical signal, and quote all subsequent results for this value of $t$.

In Fig. 2a,b we show data for $\mathcal{A}^{(i)}(\vec{x}, t = 15)$ for the pion and the rho as a function of the smearing level $i$. We find that the BS amplitude falls off less rapidly as $i$ is increased. The statistical errors are similar for smearing levels $1 - 6$ and for clarity we only show these for $i = 1$. The data show a rough convergence by $i = 6$, however, this is specific to the particular smearing method we have used and even that needs to be confirmed on a larger lattice and with further smearing levels. Given the success of this simple minded "fat" $\mathcal{M}$ in increasing the overlap with the pion state, we feel it is worthwhile to investigate other more physically motivated probes in the future. For present we take the results at $i = 6$ as our best estimate for a gauge-invariant BS amplitude.

One advantage of measuring the gauge-invariant BS amplitude is that we can follow the signal out to $x = 15$ on a $L = 16$ lattice. The data for the pion (Fig. 3) show that for $\mathcal{A}_\pi^{(1-6)}$ the large $x$ behavior is fit well by an exponential for $6 \leq x \leq 12$, while such a behavior is hard to extract from $\mathcal{A}_\pi^{(0)}$. The data for $x > 12$ show some curvature which could be due to neglecting power law corrections (discussed later) or due to finite size effects induced by smearing. Our best parameterization of the asymptotic behavior is $\mathcal{A}_\pi^{(6)} \sim e^{-0.30(1)x}$ and $\mathcal{A}_\pi^{(6)} \sim e^{-0.29(1)x}$ at $\kappa = 0.154$ and $0.155$ respectively. We contrast this rate of fall-off with that obtained from density-density correlations and the pion mass in section 5. Data for the rho are analyzed in Section 6 after we formulate the dependence of the BS amplitude on the polarization axis.

A measure of lattice discretization effects can be obtained by calculating these amplitudes at non-zero momentum with the separation $\vec{x}$ taken to be parallel or perpendicular to the direction of $\vec{p}$. If the BS amplitude of a meson at rest is characterized by $\psi(x, y, z)$, then under a boost in, say, the $\hat{z}$ direction the measured amplitude should be given by $\psi(x, y, \gamma z)$ where $\gamma = [1.47, 1.66]$ is the Lorentz contraction factor at the two values of $\kappa$. We show the data for the pion at $\kappa = 0.155$ in Fig. 4. Jackknife errors are shown only for the cases $\vec{x} \| \vec{p}$ and $\vec{p} = 0$ for clarity. The data for the case $\vec{x} \perp \vec{p}$ are consistent with those for $\vec{p} = 0$ as expected. The fall-off for the case $\vec{x} \| \vec{p}$ is much faster and the signal extends only to $x = 6$ at either value of $\kappa$. This range is not long enough to extract an effective $\gamma$; all we can say at present is that qualitatively it is consistent with the expected Lorentz contraction.

## 4. Comparison between gauge invariant, Landau, and Coulomb gauge BS amplitudes

We have measured the Coulomb and Landau gauge amplitudes for the following relative separations: the anti-quark's position is varied in a cube of size $L/2$ with respect to the position of the quark which is taken to be at $(0,0,0)$. For each of these relative separations we sum the quark's position over the time slice to produce a zero-momentum state. We find that the data for $x > 6$ along any of the axes show that there is a significant contamination from wrap-around effects due to periodic boundary conditions. These effects can be taken into account by subtracting the contributions of all the mirror points. In this paper we only use the Landau and Coulomb gauge data for comparison; for this purpose it suffices to choose data with $|x_i \leq 6|$ to avoid significant finite size effects.

The data show that $\mathcal{A}^{(6)}$ is approximately the same as $\mathcal{L}$ and slightly broader than $\mathcal{C}$. On the other hand we find $\mathcal{L} \gtrsim \mathcal{C} > \mathcal{A}^{(0)}$ consistent with the earlier results of Ref. [5]. These two features are illustrated in Fig. 5a by data for $\mathcal{C}_\pi$, $\mathcal{A}_\pi^{(0)}$ and $\mathcal{A}_\pi^{(6)}$. Similarly, in Fig. 5b we show $\mathcal{L}_\pi$, $\mathcal{A}_\pi^{(0)}$ and $\mathcal{A}_\pi^{(6)}$. The slight increase in $\mathcal{L}_\pi$ and $\mathcal{C}_\pi$ at large $x$ is most likely due to mirror contributions that arise due to the use of periodic boundary conditions.

Using six levels of smearing on a $16^3$ lattice with periodic boundary conditions produces a linear combination of paths in $\mathcal{M}$ that corresponds to a fairly smooth distribution of gauge fields across the complete time-slice. It is therefore not surprising that $\mathcal{C}$, $\mathcal{L}$ and $A^{(6)}$ give very similar results. It is still advantageous to work with gauge invariant BS amplitudes as one can probe larger separations (up to $x = 15$) without having to worry about contributions of mirror sources, and even more importantly because there are no ambiguities associated with Gribov copies.

## 5. Comparison with density-density correlations

The density-density correlation for a charged pion is defined by

$$\rho^{\alpha\beta}(\vec{x}) = \int d^3y \, \langle 0|J_\pi(T)\overline{u}(\vec{y},t)\Gamma_\alpha u(\vec{y},t)\overline{d}(\vec{x}+\vec{y},t)\Gamma_\beta d(\vec{x}+\vec{y},t)\overline{J}_\pi(0)|0\rangle \tag{5.1}$$

where $J_\pi$ is a pion source and $\overline{u}(y,t)\Gamma_\alpha u(y,t)$ probes the $u$ quark at space-time point $(y,t)$. In particular, the correlation between charge density operators, $\Gamma_\alpha = \Gamma_\beta = \gamma_4$, is a measure of the charge distribution as a function of separation between quarks [9][10]. The charge radius measured in experiments is defined as

$$\langle r_{exp}^2 \rangle = -6 \left.\frac{\partial F(q^2)}{\partial q^2}\right|_{q^2=0}, \tag{5.2}$$

where $F(q^2)$ is the pion form factor. It was shown in Ref. [14] that the Fourier transform of $\rho^{44}(x)$ is

$$\tilde{\rho}^{44}(q) = \sum_n \frac{1}{4E_n m_\pi} \langle \pi(\vec{p}=0)|J_4(\vec{x}=0)|n(q)\rangle \langle n(q)|J_4(\vec{x}=0)|\pi(\vec{p}=0)\rangle \tag{5.3}$$

where $n(q)$ is a complete set of states. Assuming that only $n = \pi$ intermediate state contributes, one can show that

$$\left.\frac{\partial \tilde{\rho}^{44}(q)}{\partial q^2}\right|_{q^2=0} = 2 \left.\frac{\partial F(q^2)}{\partial q^2}\right|_{q^2=0}. \tag{5.4}$$

However, due to the contributions of, say, excited P-wave states there is no simple relation between the charge radius and $\rho^{44}$ [15] [16]. It is not known whether these corrections enhance or decrease the rms radius calculated from $\rho^{44}$ with respect to the experimental value of the charge radius. Using the data for $\rho^{44}$ presented below we find a value that is too large, but we remind the reader that the lattice calculation has a number of systematic errors that are not accounted for.

We calculate $\rho^{44}(x)$ using Wuppertal source quark propagators. The meson sources are at time $T = 0$ and $T = 40$ corresponding to forward and backward propagation on our lattice. Even though we have measured $\rho^{44}(x)$ for both the pion and the rho, we only present results for the pion as the signal for the rho is too noisy. To improve the signal we present results for the ratio $\mathcal{R}^{\alpha\beta}(x) = \rho^{\alpha\beta}(\vec{x})/\rho^{\alpha\beta}(0)$. By dividing by the correlation function at zero separation we have also removed uncertainties due to renormalization of operators on the lattice and can concentrate on effects which depend on the separation between the two density operators.

The Wick contractions for the density-density correlation yield the three types of diagrams shown in Fig. 6. Fig. 6a shows both insertions on valence quarks, Fig. 6b shows the correlation between insertion on a sea quark loop and a valence quark within the pion, and Fig. 6c shows the correlation between two sea quarks within the pion. All three diagrams can be estimated in the quenched approximation, however, in order to include diagrams in Figs. 6b,c we require additional calculation of quark propagators from every spatial point. This is beyond the scope of the present calculation. Furthermore, each of these three diagrams is modified by sea quark contributions; these remain a source of systematic error that we cannot yet estimate.

Hueristically, we expect the effects of diagrams shown in Figs. 6b,c to be suppressed. The loop $\overline{\psi}\gamma_4\psi$ can interact through gluon exchange with either a valence quark in the pion or with a second loop. At short distances, and when the pion propagates as shown schematically in Fig. 7a (with the density insertions replaced by disconnected loops), these interactions are suppressed by powers of $\alpha_s$. On the other hand for cases corresponding to propagation shown in Fig. 6c the exchange is suppressed by $e^{-m_g x}/e^{-m_\pi x}$ where $m_g$ is the mass ($\sim 1\ GeV$) of the mixed glueball meson state with the appropriate quantum numbers. Needless to say, it is primarily for computational reasons that we restrict ourselves to the connected diagram of Fig. 6a which can be labeled as the valence density correlation in the quenched approximation.

The asymptotic behavior of the connected density correlation is governed by amplitudes which fall most slowly as a function of spatial separation $x$. In our setup, the time separation between the source (sink) and the density probes is $> 10a$, so, based on our analysis of the spectrum and the BSA, we are probing the lowest state in the pion channel. This $t$ separation is also larger than the maximum spatial separation allowed between the density probes, i.e. $x_i = 8$, so we neglect transient effects due to the spatial position and character of the pion source (sink). The leading contributions are then described schematically by the two diagrams shown in Figs. 7a,b. The diagram in Fig. 7a shows that the pion can propagate by preserving its identity all the way, so $\mathcal{R}^{44}(x)$ should fall off as $x^{-n}e^{-m_\pi x}$ as $x \to \infty$. The factor $x^{-n}$ comes from the superposition of momentum excitations in the propagation of any state created by a localized source, and in three dimensions the asymptotic form is $x^{-3/2}$. In Fig. 7b only one of the quarks propagates in the spatial direction, effectively forming a flavor singlet meson with a $\gamma_i$ point source (under a rotation of axis $\gamma_4 \to \gamma_i$). Thus the fall off is expected to go as $x^{-3/2} e^{-m_\rho x}$ for large $x$. So for pions the scenario in Fig. 7a produces the dominant behavior at large distances, while for the rho both diagrams fall off at the same exponential rate, i.e. $e^{-m_\rho x}$.

We measure $\rho_\pi^{44}(x)$ at all $\vec{x}$ in the first Brillouin zone. Nevertheless, we find that it is necessary to account for the leakage between adjacent Brillouin zones due to the spatial periodicity of the lattice [17] [18]. The lattice correlation, for a cubic lattice of side $L$, denoted by $\rho(\vec{x})_L$, is an infinite sum of all the mirror images

$$\rho(\vec{x})_L = \sum_{n_1,n_2,n_3=\infty}^{\infty} \rho(\vec{x} + L\vec{n})_\infty . \tag{5.5}$$

In practice we truncate the sum and consider contributions only from $|x_i - n_i| \leq L$. This truncation is justified since the density correlation falls exponentially at large separations with a decay rate equal to the pion mass, thus the contribution from the nearest images is roughly $e^{-m_\pi L} \approx e^{-5.6}$ with a maximum degeneracy factor of 8. To perform the correction for the periodic images we parametrize $\rho_\pi^{44}(x) \sim e^{-m_\pi x}$ (we ignore the additional power law factor as our data are not sensitive to it) self-consistently until it gives the best fit to the data. Further details of this method for image corrections are given in a forthcoming paper[17].

The raw data for the density correlation and the Bethe-Salpeter amplitude are shown in Fig. 8. To interpret these results we first define and extract an rms radius from each quantity and then discuss the physical meanings of these. Assuming Eq. (5.4), the charge radius is given in terms of the density-density correlation by

$$\langle r_{dd}^2 \rangle = \frac{1}{2} \int d^3x x^2 \rho^{44}(x) \tag{5.6}$$

where $x$ is the relative separation of the $\overline{d}$ and $u$. Similarly, we define an rms radius from the BSA amplitude as

$$\langle r_{BSA}^2 \rangle = \frac{1}{w} \frac{\int d^3x x^2 (\mathcal{A}^{(6)}(x))^2}{\int d^3x (\mathcal{A}^{(6)}(x))^2} \tag{5.7}$$

where the factor $w$ translates the relative separation to separation from the center of mass. The two limiting cases are $w = 4$ for a two body system with degenerate $q$ and $\overline{q}$, and $w = 2$ if the $q$ and $\overline{q}$ move independently about the center of mass. The real situation is somewhere in between, and at present we do not have a way of estimating it [15]. We shall calculate $r_{BSA}$ assuming $w = 4$ and comment on this uncertainty later.

To calculate these rms radii from our data requires that we make an ansatz for the asymptotic fall-off. For the density-density correlations we use both $e^{-m_\pi x}$ and $x^{-3/2} e^{-m_\pi x}$ where $m_\pi$ is the pion mass extracted from the 2-point function. We regard the difference between the results from the two ansätze a measure of the uncertainty induced by this process. For the BSA we simply use the fits shown in Fig. 3.

The final results are given in Table 1, and these are to be compared with the experimental value $r_\pi = 0.636 \pm 0.037$ fm [19]. We also include for comparison results on $16^4$ lattices at $\beta = 5.7$. A description of this data set is given in Ref. [18], and we have reanalyzed it as described in this paper. A comparison of $r_{dd}$ obtained at the two values of $\beta$ is particularly significant because the pion mass is roughly the same in physical units. This is true at each of the two values of the quark mass. The data show a significant variation with the quark mass and extrapolating to the physical pion mass will give an $r_{dd}$ that is significantly larger than the experimental value at both values of $\beta$. Chiral perturbation theory predicts that $r_\pi$ has a logarithmic singularity in the chiral limit [20]. In the quenched approximation, however, the pion loop that gives rise to this singularity is absent, so we expect a finite value for the rms radius in the chiral limit. The data also show that the deviation is significantly reduced on going from $\beta = 5.7 \to 6.0$; to get the continuum result one needs to work at still weaker coupling or to use an improved action. Also, the effects of the disconnected graphs, *i.e.* Figs. 6b,c, and quenching need to be investigated before we can quantify the effects of the neglected $(n \neq \pi)$ states in Eqs. (5.3) (5.4).

In contrast to density-density correlations, $r_{BSA}$ is not very sensitive to the quark mass over the small range investigated here and it is roughly a factor of two smaller than the experimental value. As discussed above, using a more realistic value for $w$ will reduce this deviation. Nevertheless, based on the argument given below we expect the rms radius measured from the BSA to be smaller than the experimental value.

The pion wavefunction on any given spacelike surface can be schematically decomposed into its Fock space components made up of quarks and gluons

$$|\pi\rangle = |\overline{d}\gamma_5 u\rangle + c_n^0 |\overline{d}\gamma_5 u(\mathrm{n}g)\rangle + c_n^1 |\overline{d}\gamma_5 u\overline{q}q(\mathrm{n}g)\rangle + \ldots \qquad (5.8)$$

On the lattice the higher Fock states are present because the quark propagator, even in the quenched approximation, has an overlap with multi-quark and gluon states due to the back and forth propagation across any spacelike surface. The BSA only probes those Fock states in which $\overline{d}\gamma_5 u$ is traced over spin and color to match the quantum numbers of a pion, while the density-density correlation has no such restriction. This distinction is crucial because the charge radius gets enlarged if the strong force between the $\overline{d}$ and $u$ can be screened by popping $q\overline{q}$ pairs from the vacuum. This is shown schematically in Fig. 9 for the density-density correlation with both the $u$ and $\overline{d}$ forming a color singlet state with a $q\overline{q}$ pair produced either by a "Z excursion" or by vacuum polarization. Due to the restriction of spin and color trace on $\overline{d}\gamma_5 u$ only those higher Fock states which factorize, for example $|\overline{d}\gamma_5 u \overline{q}q(ng)\rangle \to |\overline{d}\gamma_5 u\rangle |\overline{q}q(ng)\rangle$ or those that can be rearranged into this form by spin and color Fierz, will contribute to the BSA. For this reason the rms radius calculated from BSA is expected to be an underestimate of the experimental value.

Using the same set of propagators and image correction scheme, we have measured the pseudoscalar correlation $\mathcal{R}^{55}(x)$ for the pion. The radial distribution $x^2 \mathcal{R}^{55}_\pi(x)$ at $\kappa = 0.154$ is plotted in Fig. 10. The asymptotic behavior of $\mathcal{R}^{55}(x)$ is simpler to analyze than $\mathcal{R}^{44}(x)$; since the $\gamma_5$ insertions are unchanged under a Wick rotation the exponential decay of $\mathcal{R}^{55}(x)$ for large $x$ is controlled by the pion mass. The data in Fig. 10 show that a $16^3$ lattice is too small to observe this asymptotic decay. In Ref. [18] it was shown on $16^4$ lattices at $\beta = 5.7$, corresponding to a physical volume larger by a factor of two, that the asymptotic behavior is consistent with the expected behavior

The volume integral of $\rho^{55}$ characterizes interactions between quarks. For example in the MIT Bag model, Lissia *et al.* [21] have shown that for hadronic systems constructed from noninteracting quarks, *i.e.* where the composite wavefunction is a direct product of S-wave single quark wavefunctions that satisfy the Dirac equation, the volume integral $\int d^3x x^2 \rho^{55}_\pi(x)$ is zero. Deviations of this volume integral from zero result from interactions between the quarks. The present data show the requisite node but we cannot estimate the integral with sufficient accuracy due to the finite size cutoff, so we cannot test the above hypothesis.

## 6. Polarization dependence of the rho meson BSA

The $\rho$ meson wavefunction is a linear combination of $l = 0$ and $l = 2$ orbital angular momentum states. On the lattice, where rotational invariance is absent, the BS amplitude decomposes into irreducible representations of the cubic group:

$$\langle 0|\overline{d}(\vec{x})\gamma_i \mathcal{M}(\vec{0}, \vec{x}) u(\vec{0})|\rho(\vec{0}, j)\rangle$$
$$= \frac{m_\rho^2}{f_\rho} \left( \delta_{ij} \phi_{A_1}(\vec{x}) + \delta_{ij} \left( \frac{x_i^2}{\vec{x}^2} - \frac{1}{3} \right) \phi_E(\vec{x}) + (1 - \delta_{ij}) \frac{x_i x_j}{\vec{x}^2} \phi_{T_2}(\vec{x}) \right) \quad (6.1)$$

where $|\rho(\vec{0}, j)\rangle$ is a state of momentum $\vec{0}$ and polarization $j$. The functions $\phi_{A_1}$, $\phi_E$ and $\phi_{T_2}$ are scalars which multiply tensors transforming under the cubic group as the $A_1$, $E$ and $T_2$ representations (of dimensions 1, 2 and 3 respectively). As is clear from the tensor structure, $A_1$ corresponds to the $l = 0$ state while $E$ and $T_2$ together form the decomposition of the $l = 2$ state.

Lattice calculations therefore allow us to investigate, as a function of the quark mass, the relative mixture of $l = 0$ and $l = 2$ states ($A_1$ versus $E$ or $T_2$), and the restoration of rotational symmetry (degeneracy of $E$ and $T_2$) by studying the three cases: (A) $i = j$ and $\vec{x}$ along $i$ ($\|$), (B) $i = j$ and $\vec{x}$ perpendicular to $i$ ($\perp$), and (C) $i \neq j$.

At present we only measure the BS amplitude for $i = j$, and the results for cases (A) and (B) at $\kappa = 0.154$ are shown in Fig. 11. The data show that for large separation the fall-off is extremely well fit by an exponential in both cases, with a rate of fall-off given by $m_\| = 0.226(8)$ and $m_\perp = 0.263(7)$ respectively. It is interesting to note that the large $x$ fall-off is governed by a mass that is roughly $m_\rho/2 = 0.23$. Such a behavior would be expected if the rms radii defined in Eqns. (5.6) and (5.7) were the same.

From the data shown in Fig. 11 we extract $\phi_{A_1}(x)$ and $\phi_E(x)$. The results, shown in Fig. 12, are fit to simple hydrogen-like radial wavefunctions

$$\begin{aligned} \phi_{A_1}(x) &= 1.61 \, e^{-0.242 x} , \\ \phi_E(x) &= 0.029 \, x^2 \, e^{-0.375 x} . \end{aligned} \quad (6.2)$$

These functions give a good fit to $\phi_{A_1}(x)$ for $x \geq 6$ and $\phi_E(x)$ for $x \geq 2$. The results at $\kappa = 0.155$ are qualitatively similar. The data for both $\phi_{A_1}(x)$ and $\phi_E(x)$ are slightly broader, though the difference is smaller than the statistical errors.

## 7. Conclusions

We show how to define gauge invariant BSA using smeared links that have similar overlap with meson wavefunctions as those in Coulomb or Landau Gauge. We use the gauge invariant BSA to calculate the rms radius for the pion and qualitatively demonstrate the Lorentz contraction along the direction of motion for a pion.

For the rho meson the wavefunction is characterized by three different orbital functions $\phi_{A_1}$, $\phi_E$ and $\phi_{T_2}$. These correspond to the $A_1$, $E$ and $T_2$ representations of the cubic group, which in the continuum limit become the $l = 0$ and $l = 2$ representations. We show how each of these three functions can be extracted by choosing the polarization and separation axis appropriately. A test of restoration of rotational invariance can be made by comparing $\phi_E$ and $\phi_{A_1}$. With the present data we can only calculate $\phi_{A_1}$ and $\phi_E$ and show that these are well described by simple $l = 0$ and $l = 2$ hydrogen-like wavefunctions respectively.

The rms radius of the pion measured from the BSA is roughly $0.5 - 0.7$ of the experimentally measured value. The large uncertainty is due to the lack of knowledge of the position of the center of mass with respect to the valence $\overline{d}$ and $u$ due to the motion of the glue and additional $q\overline{q}$ pairs. We attribute the remaining discrepancy to the following two factors. First, the BSA does not fully probe those parts of the wavefunction in which there is a saturation of the color force between the valence quarks and $q\overline{q}$ pairs in the sea. Second, in the quenched approximation, sea quark contributions of the type shown in Fig. 9b are absent.

Present data show that the pion's radius extracted from density-density correlations has a significant variation with the quark mass. An extrapolation to the physical pion mass gives a significantly larger value than the experimental number. This is not a cause for concern as $r_{dd}$ is not simply related to the experimentally measured charge radius. Furthermore a comparison of data between $\beta = 5.7$ and $6.0$ shows that the deviation decreases with the lattice spacing and suggests that taking the continuum limit would give a result much closer to the experimental value. Lastly, one needs to evaluate the effects of using the quenched approximation, excited state contamination, and the disconnected diagrams that have been neglected in this calculation.

## Acknowledgments

These calculations were done on Cray YMP using time provided by DOE at NERSC, by NSF at Pittsburgh and San Diego Supercomputer centers and by a Cray Research grant at PSC. We are very grateful to Jeff Mandula and Ralph Roskies for their support of this project. We also thank Matthias Burkardt, Peter Lepage, Jeff Mandula, John Negele and Steve Sharpe for informative discussions.

| $\beta$ | $\kappa$ | $m_\pi$ | $r_{dd}[e^{-m_\pi x}]$ | $r_{dd}[x^{-3/2}e^{-m_\pi x}]$ | $r_{BSA}[e^{-mx}]$ |
|---|---|---|---|---|---|
| 6.0 | 0.154 | 0.365a | 6.8(1)a | 6.5(1)a | 3.01a |
| | | 700MeV | .72(1)fm | .68(1)fm | 0.32fm |
| 6.0 | 0.155 | 0.296a | 7.6(1)a | 7.0(2)a | 3.08a |
| | | 560MeV | .80(1)fm | .74(2)fm | 0.33fm |
| 5.7 | 0.160 | 0.694a | 3.8(1)a | 3.8(1)a | |
| | | 694MeV | .76(1)fm | .76(1)fm | |
| 5.7 | 0.164 | 0.527a | 4.5(1)a | 4.5(1)a | |
| | | 527MeV | .89(1)fm | .89(1)fm | |

**Table 1.** Results for the pion's rms radius calculated from the BSA amplitude and the density-density correlations $\rho_\pi^{44}(x)$. The results are given in both lattice units and in physical units. We use $a = 0.2$fm and $a = 2/19$fm for the lattice scale at $\beta = 5.7$ and 6.0 respectively. The estimate for $r_{BSA}$ has 20% uncertainty. The experimental value is $r_\pi = 0.636 \pm 0.037$fm.

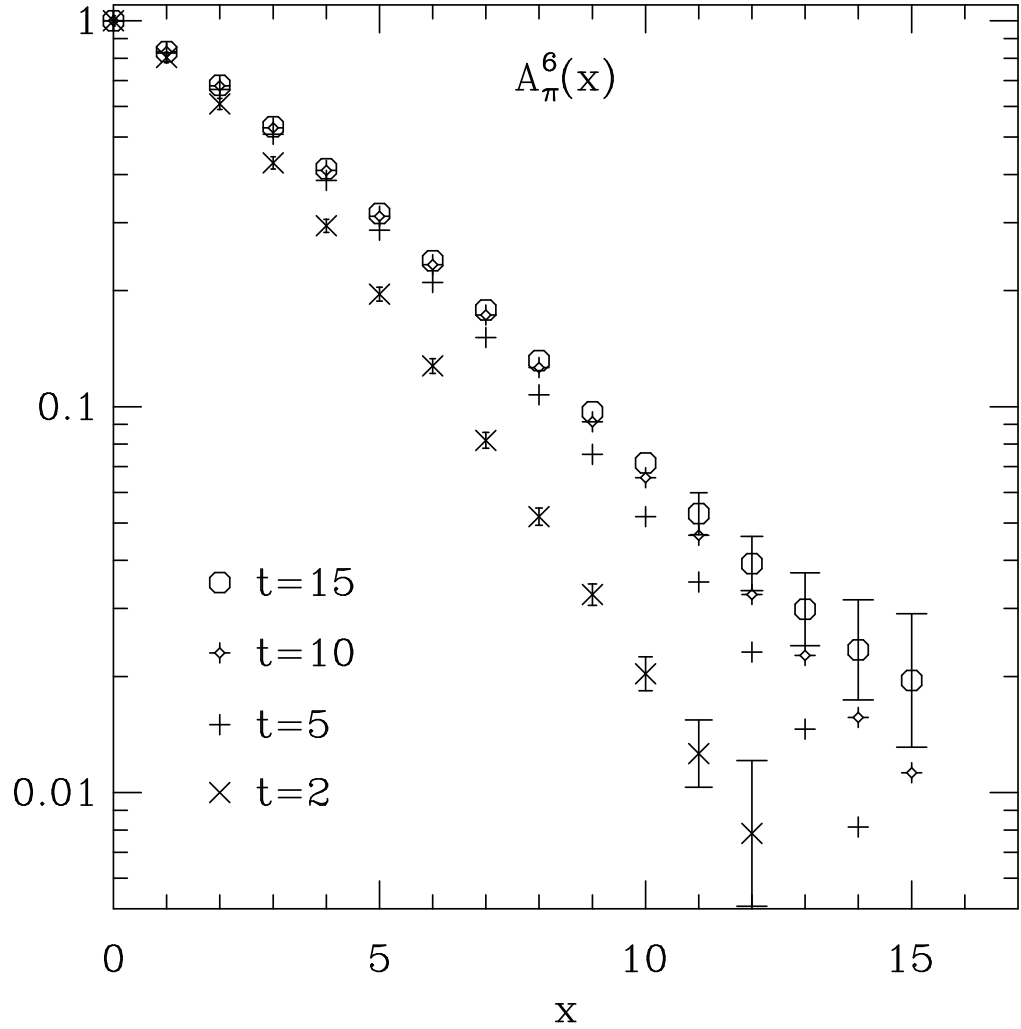

**Fig. 1a.** Gauge invariant BS amplitude for the pion at various time slices from the source. The data are obtained after the gauge links have been smeared 6 times as described in the text.

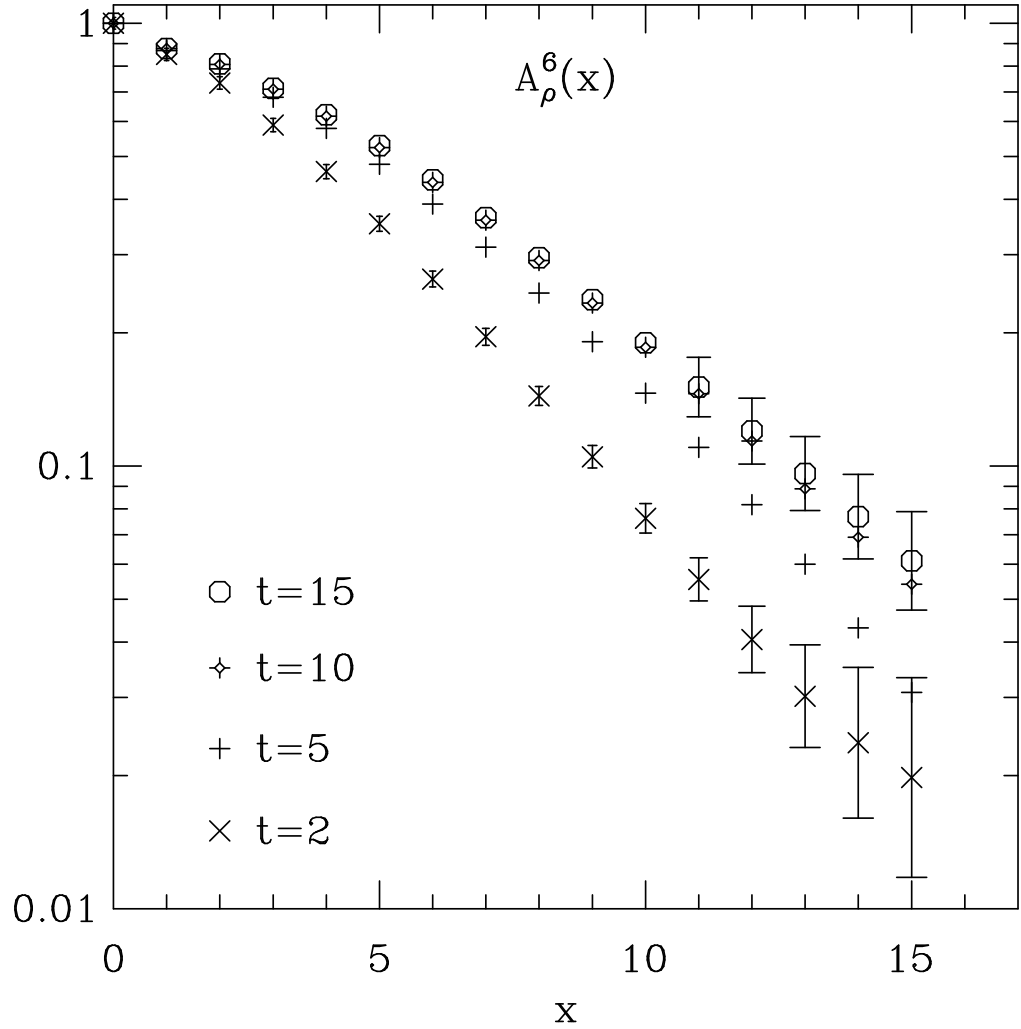

**Fig. 1b.** Same as Fig. 1a for the $\rho$ meson. The data are for the case where the separation $x$ is taken along the polarization axis.

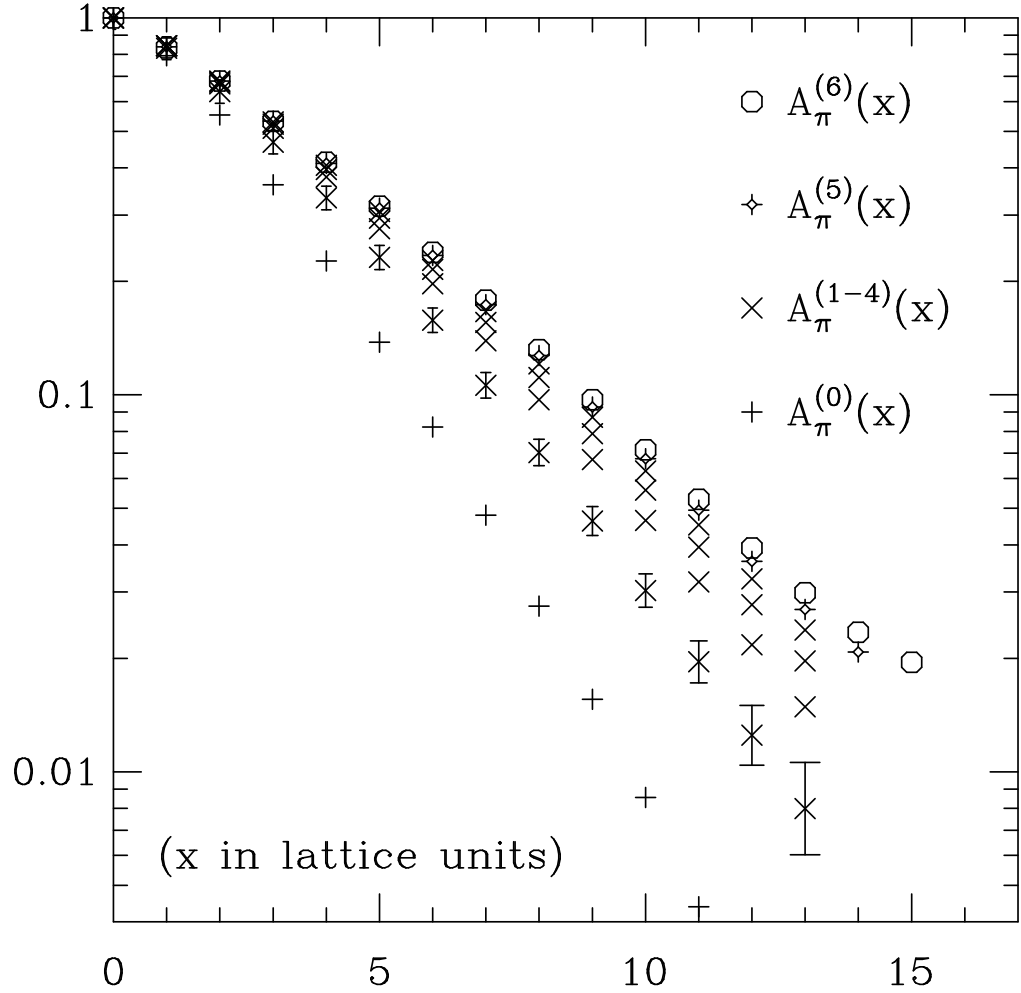

**Fig. 2a.** The gauge-invariant BS amplitude at $t = 15$ as a function of smearing level.

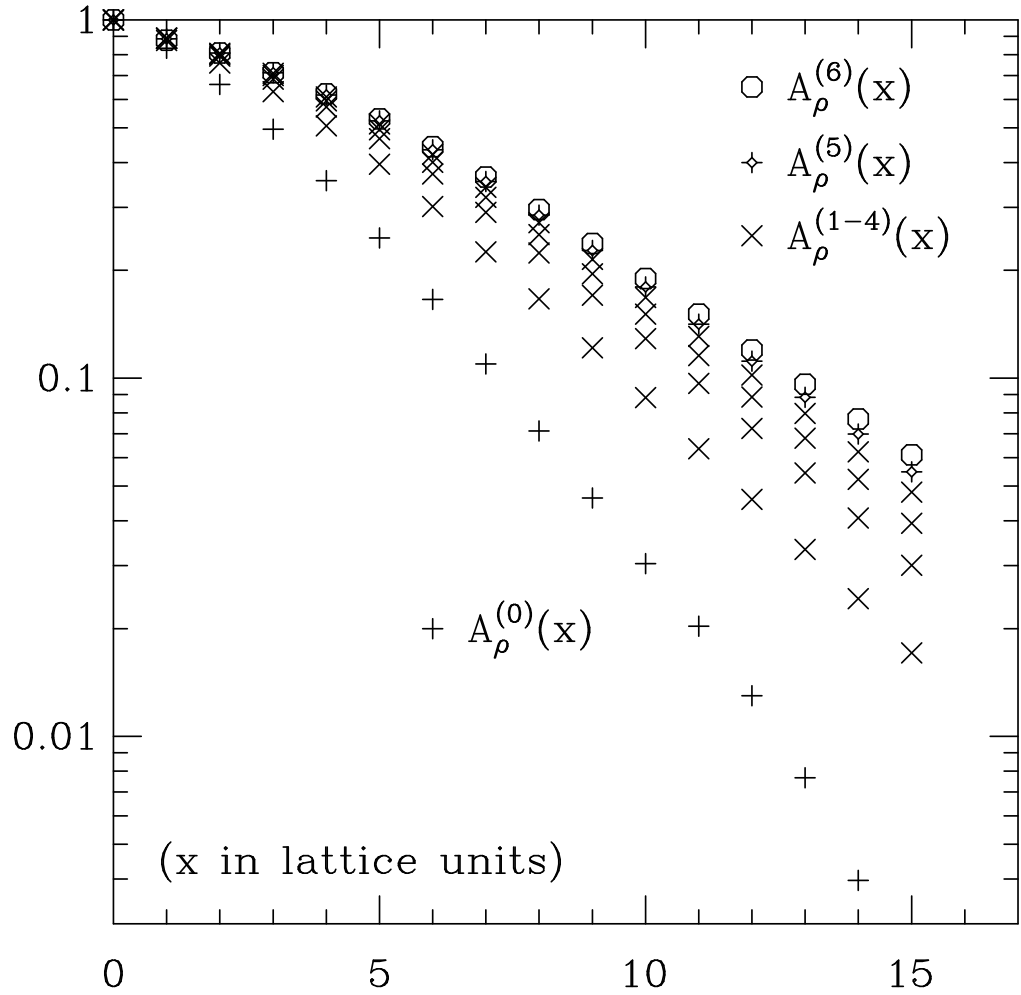

**Fig. 2b.** Same as Fig. 2a but for the $\rho$ meson.

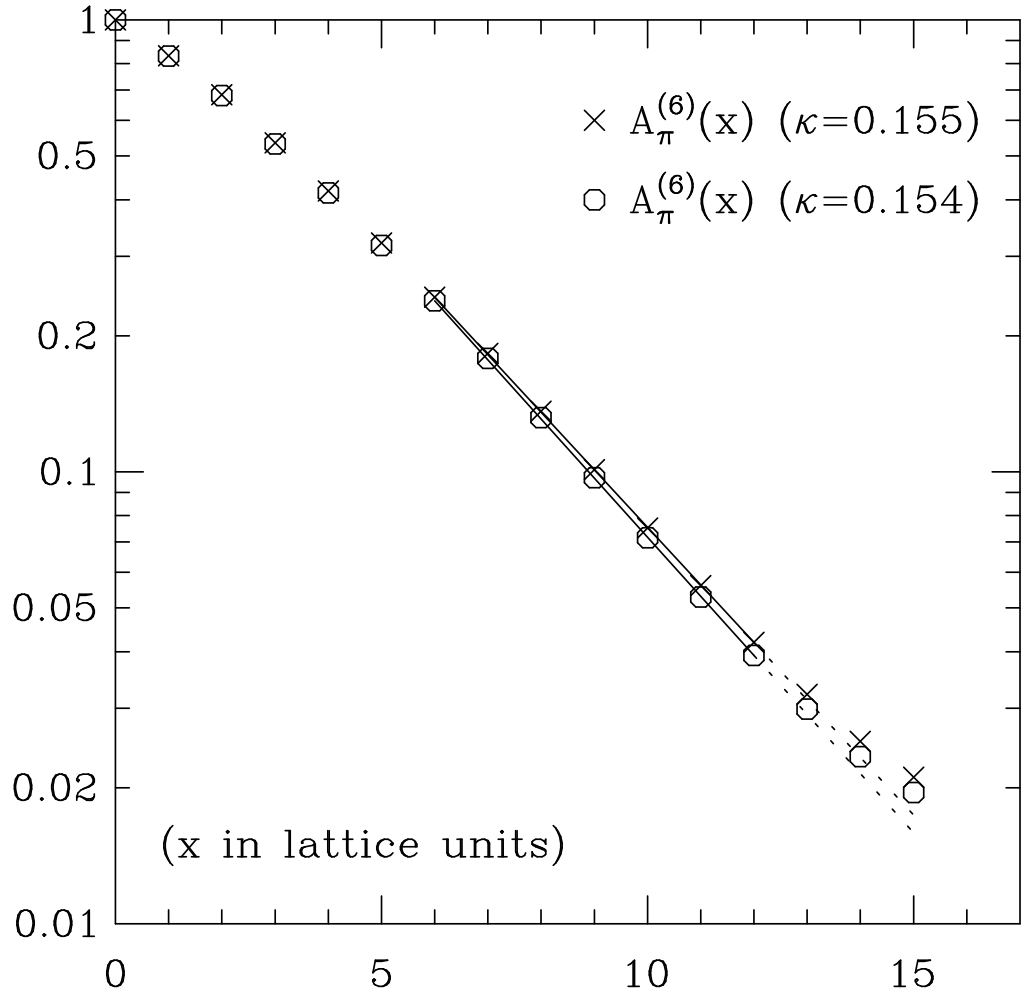

**Fig. 3.** Exponential fit to the gauge invariant BS amplitude for the pion at large $x$. The rate of fall-off is extracted from a fit to points $6 \leq x \leq 12$ at each of the two masses.

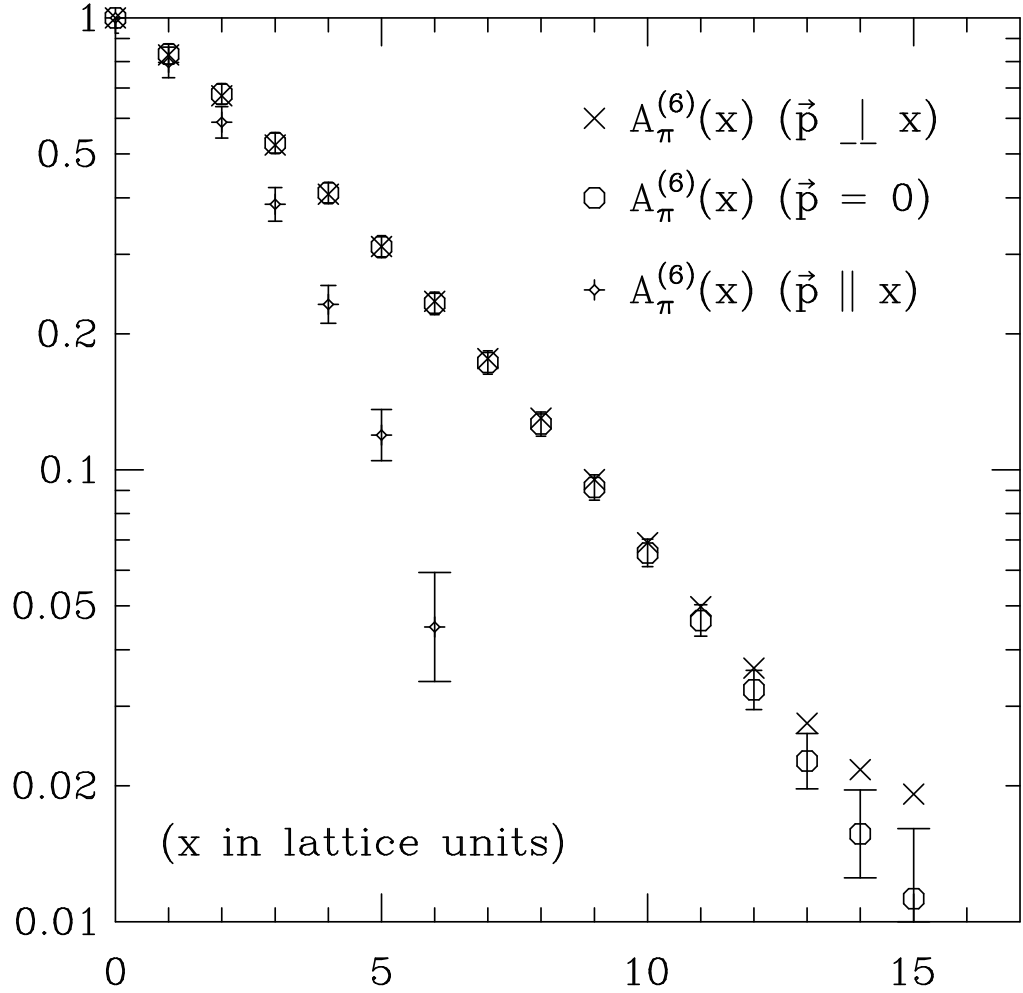

**Fig. 4.** Comparison of the pion BS amplitude at zero and non-zero momentum. The data show Lorentz contraction when $\vec{x} \| \vec{p}$ and $\vec{p}a = (0, 0, 2\pi/16)$.

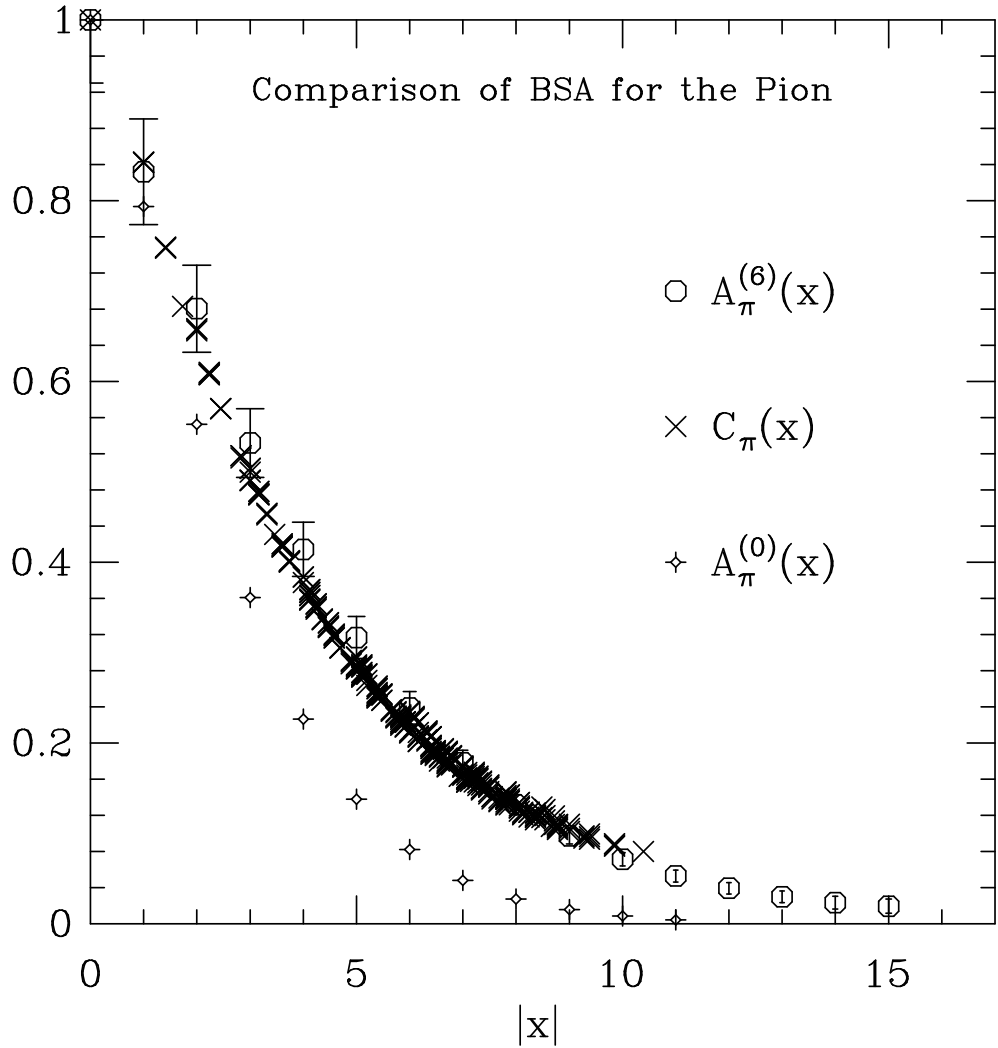

**Fig. 5a.** Comparison of $\mathcal{C}_\pi$, $\mathcal{A}_\pi^{(0)}$ and $\mathcal{A}_\pi^{(6)}$ at $\kappa = 0.154$. The data for $\mathcal{C}_\pi$ at large $x$ have not been corrected for contributions from mirror sources.

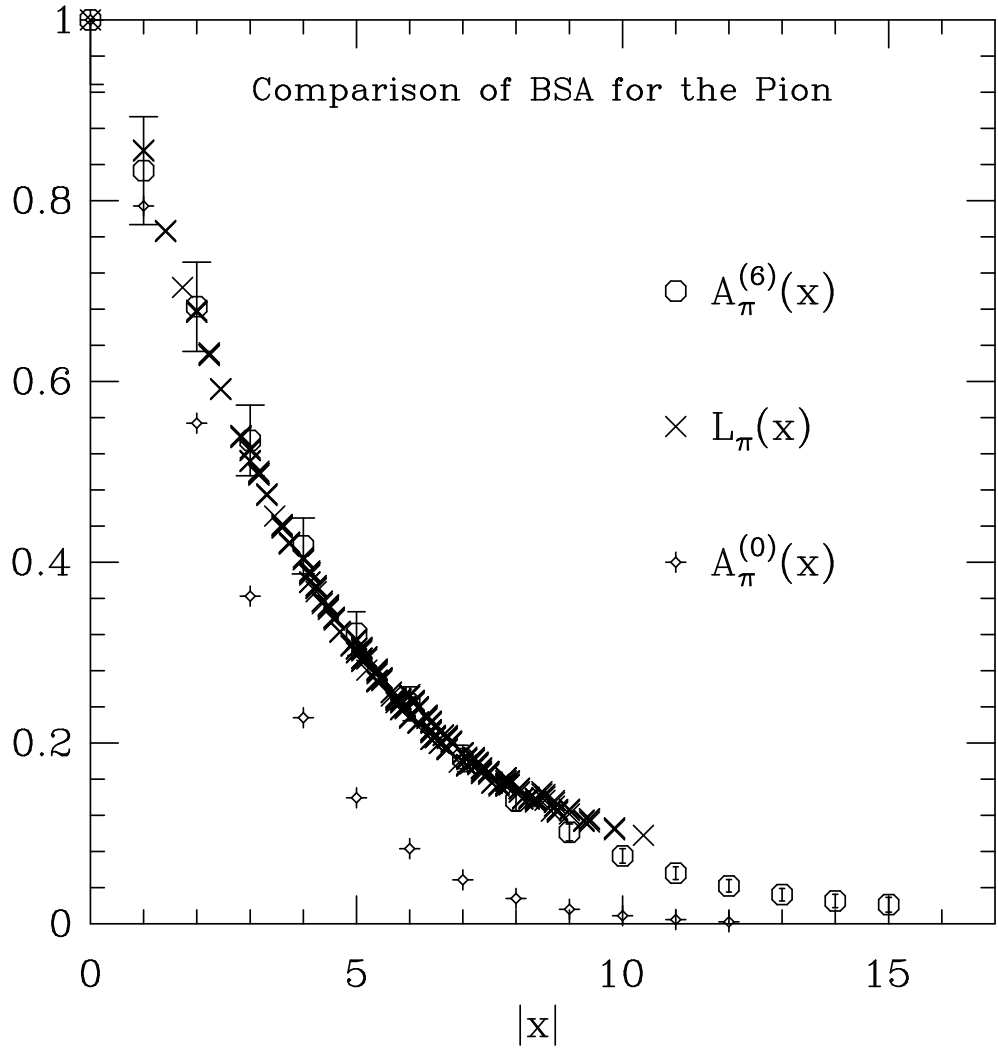

**Fig. 5b.** Comparison of $\mathcal{L}_\pi$, $\mathcal{A}_\pi^{(0)}$ and $\mathcal{A}_\pi^{(6)}$. Rest is same as Fig. 5a.

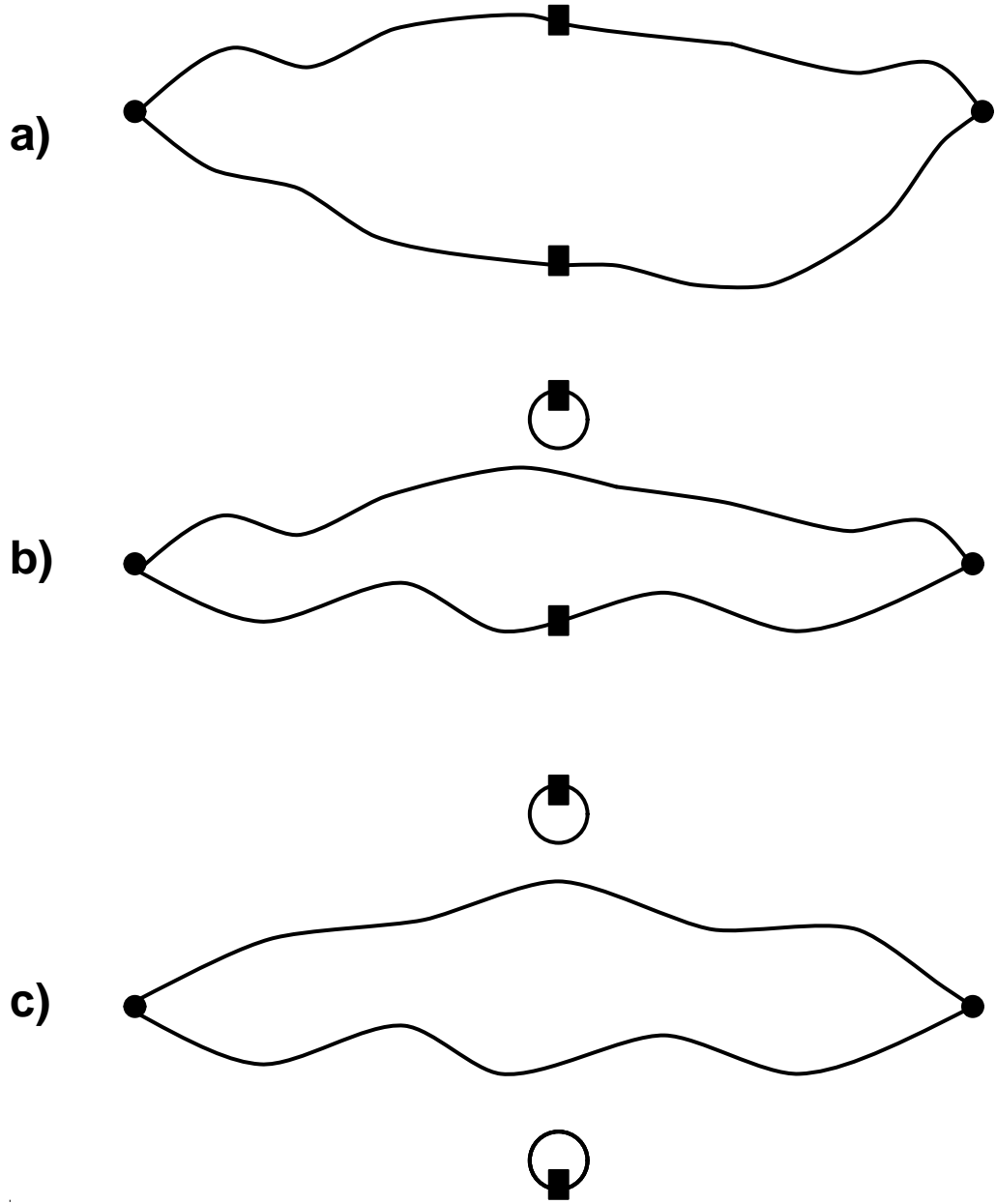

**Fig. 6** The three different Wick contractions that contribute to the density-density correlation. The filled circles denote the meson source and sink while the filled rectangles are the density insertions.

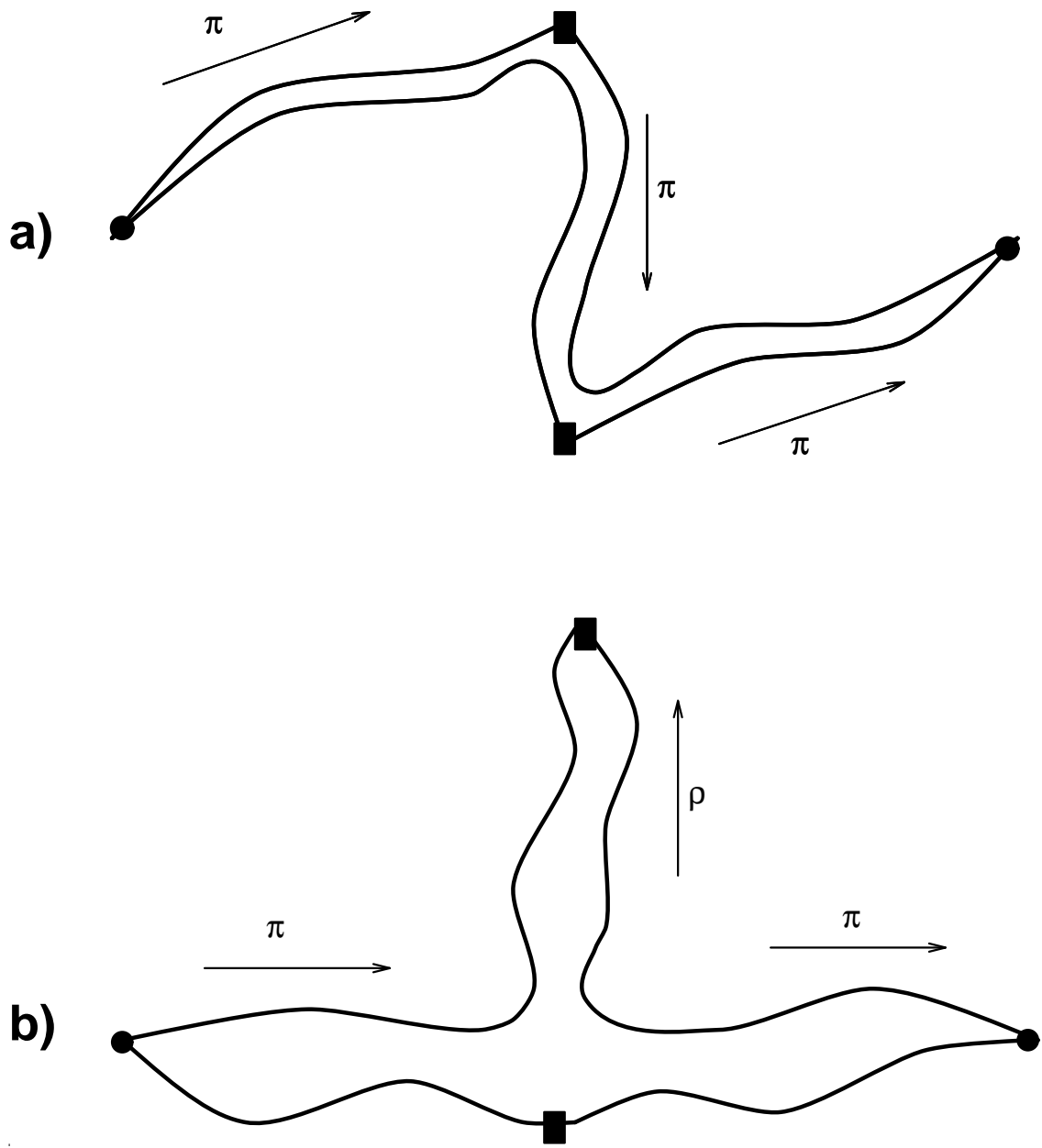

**Fig. 7** The two different types of diagrams that give the leading large $x$ behavior for the density-density correlation.

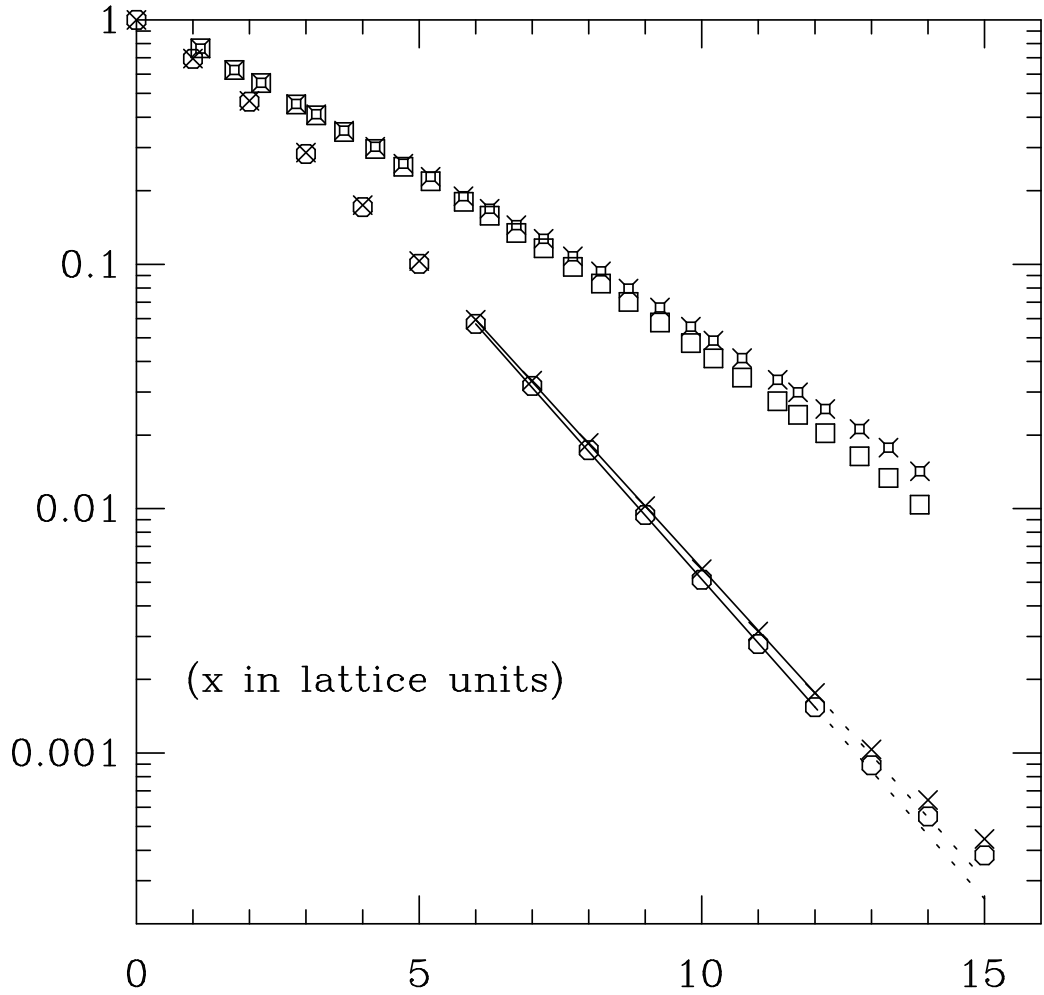

**Fig. 8** Comparison of the square of the BSA for the pion with the density-density correlation $\rho^{44}(x)$. The data are at $\kappa = 0.154$.

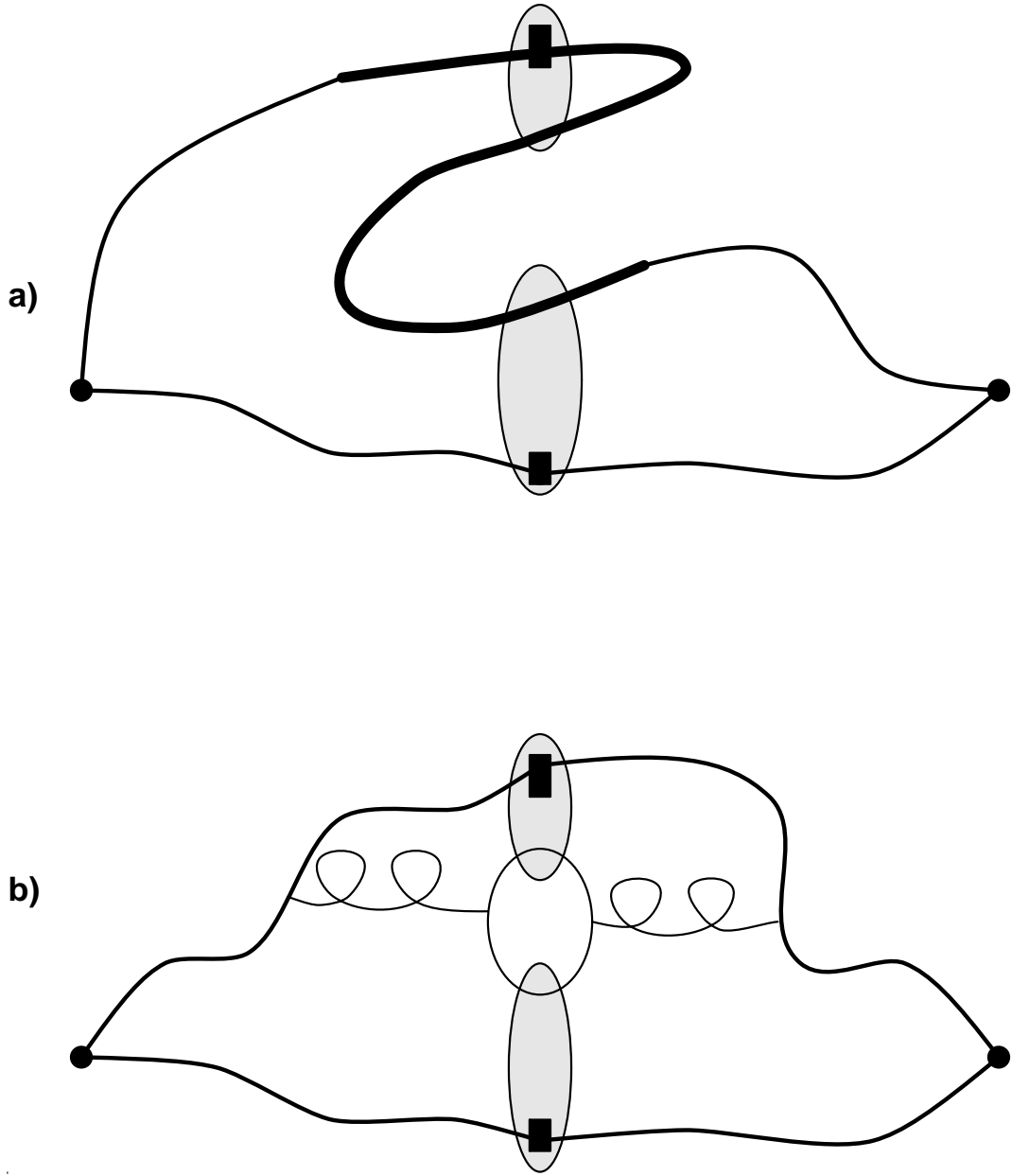

**Fig. 9** We show two example of how the color force can be saturated in the the density-density correlation. The filled rectangles are the density insertions and the shaded blobs are a schematic representation of a color neutral object. The thick line in (a) shows a "Z excursion" by the quark propagator.

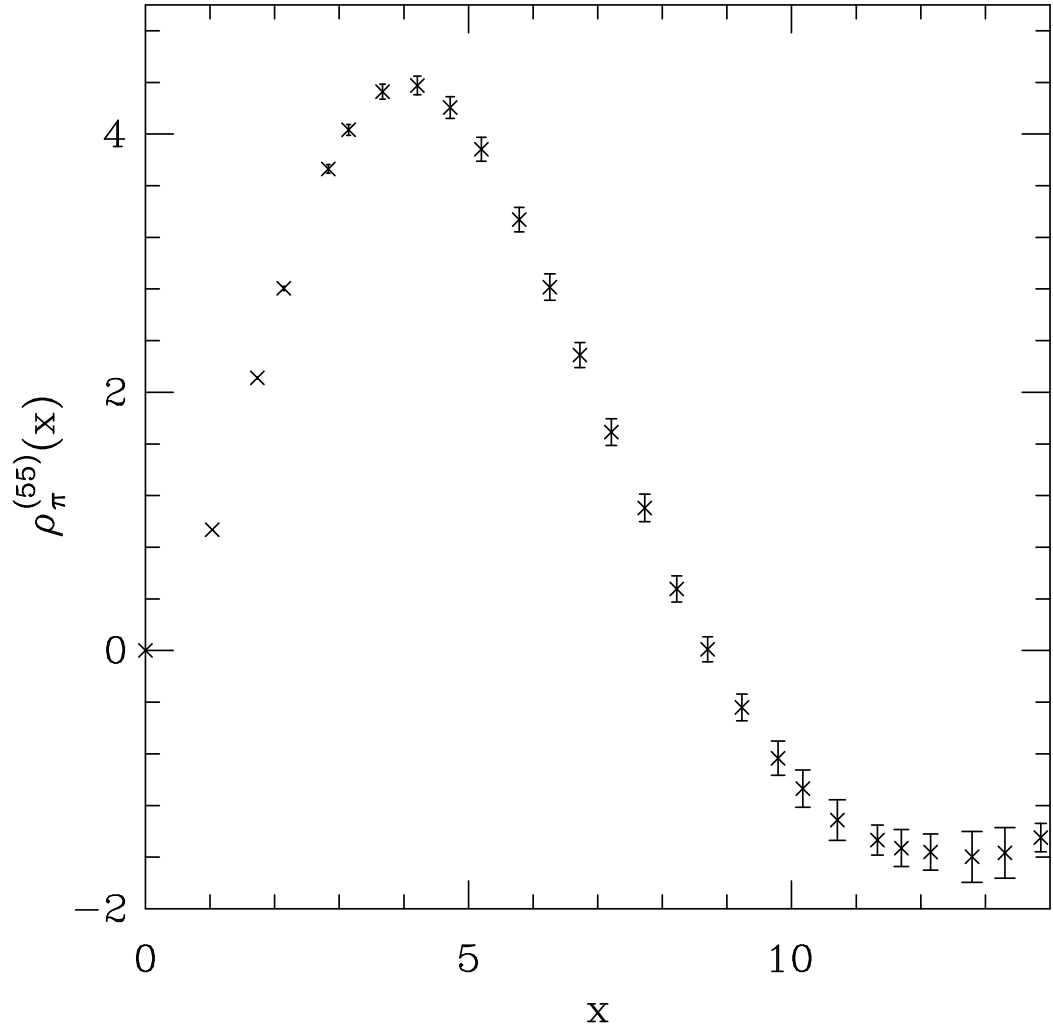

**Fig. 10** The fall-off with $x$ of the $\rho^{55}(x)$ density-density correlation inside a pion. The correlator has a node at $x \approx 8.7a$. The lattice is not large enough to verify the expected asymptotic behavior $e^{-m_\pi t}$.

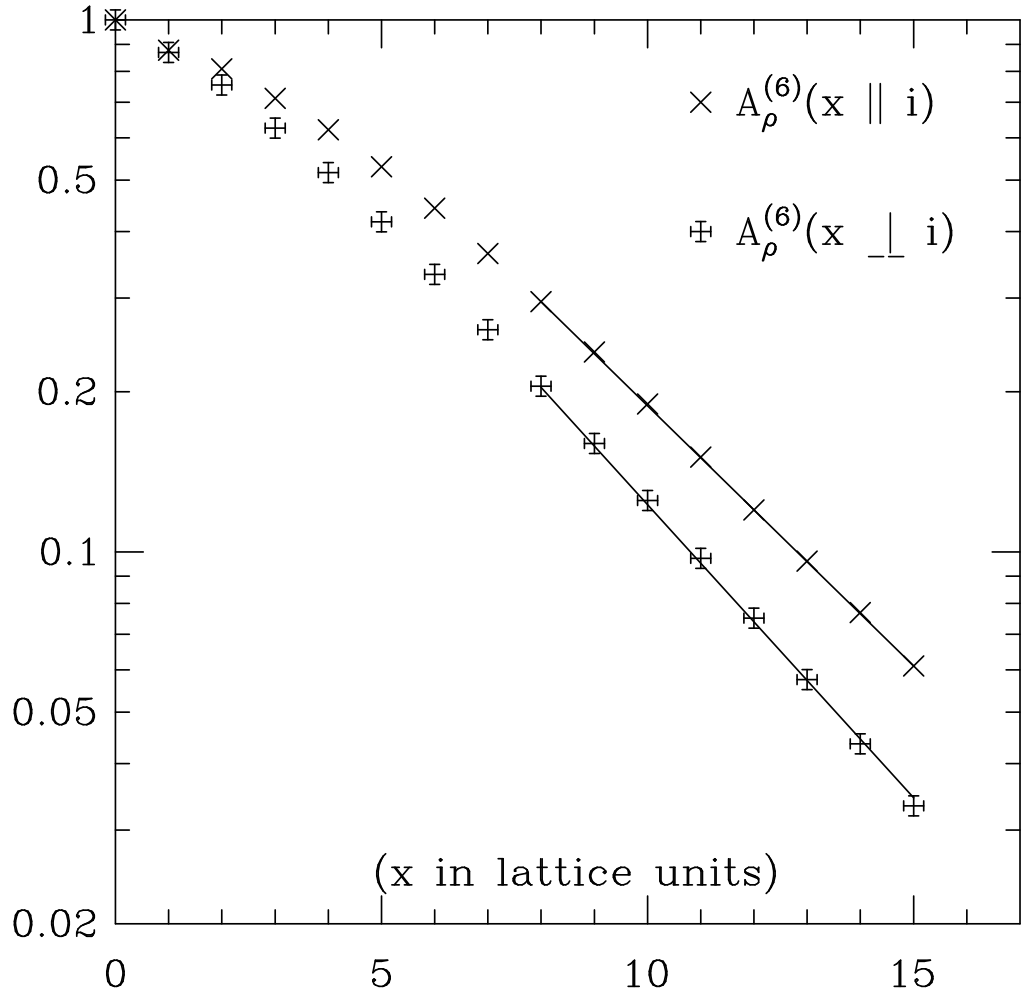

**Fig. 11** The BS amplitude for the rho at $\kappa = 0.154$ with polarization axis $i \parallel$ and $\perp$ to separation $\vec{x}$.

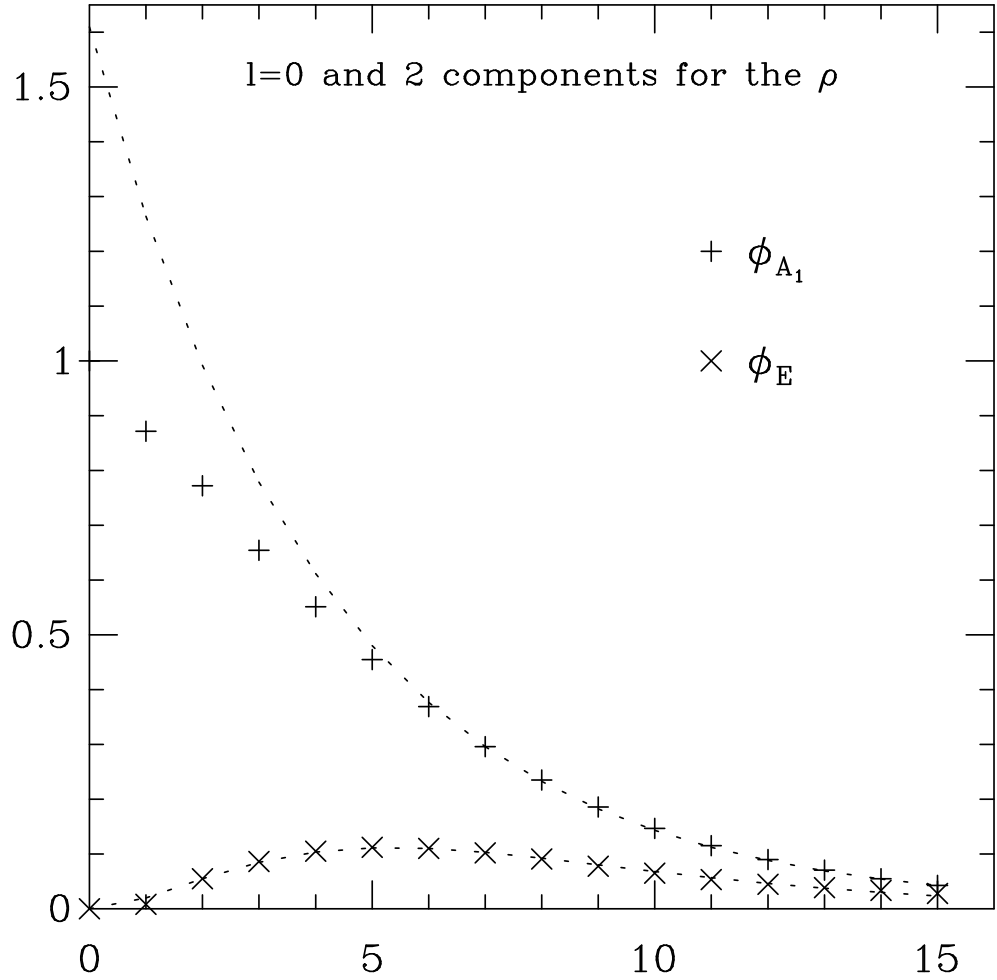

**Fig. 12.** Results for $\phi_{A_1}(x)$ and $\phi_E(x)$ using the data shown in Fig. 11. The fits are described in Eq. (6.2).